\documentclass[prb,twocolumn]{revtex4-1} 

\usepackage{amsmath}  
\usepackage{amsfonts} 
\usepackage{graphicx} 

\usepackage{enumitem}

\usepackage{hyperref}

\usepackage[table]{xcolor}
\usepackage{soul}

\usepackage{pifont}
\usepackage{newunicodechar}
\newunicodechar{✓}{\ding{51}}

\begin{document}


\title{QISCIT: A validated concept inventory assessment for quantum information science}


\author{Kelley Durkin}
\email{kelley.durkin@vanderbilt.edu}
\affiliation{Department of Teaching and Learning, Vanderbilt University, Nashville, TN 37203}

\author{Manshuo Lin}
\email{manshuo.lin@utdallas.edu}
\affiliation{Department of Physics, University of Texas at Dallas, Richardson, TX 75080}

\author{Michael H. Kolodrubetz}
\email{mkolodru@utdallas.edu}
\affiliation{Department of Physics, University of Texas at Dallas, Richardson, TX 75080}

\author{Ryan P. McMahan}
\email{rpm@vt.edu}
\affiliation{Department of Computer Science, Virginia Tech, Blacksburg, VA 24061}


\date{\today}

\begin{abstract}
Quantum information science (QIS) is a critical interdisciplinary field that requires a well-educated workforce in the near future. Numerous researchers and educators have been actively investigating how to best educate and prepare such a workforce. An open issue has been the lack of a validated tool to asses QIS understanding without requiring college-level math. In this paper, we present the systematic development and content validation of a new assessment instrument called the Quantum Information Science Concept Introductory Test (QISCIT). With feedback from 11 QIS experts, we have developed and validated a 31-item version of QISCIT that covers concepts like quantum states, quantum measurement, qubits, entanglement, coherence and decoherence, quantum gates and computing, and quantum communication. In addition to openly sharing our new concept inventory, we discuss how introductory QIS instructors can use it in their courses.
\end{abstract}

\maketitle 

\section{Introduction} 

Quantum information science (QIS), which exploits the rules of quantum mechanics to improve information processing, is expected to greatly impact the future with the rapid development of quantum computers, sensors, networks, and other technological advances. However, a well-educated QIS workforce is required to make such advances. This is complicated by the fact that QIS is a challenging interdisciplinary field to learn, involving physics, computer science, mathematics, and engineering. 

Researchers and educators have been developing and investigating a wide variety of curricula, frameworks, tools, and technologies to facilitate and enhance QIS education. In a recent workshop funded by the \textit{National Science Foundation (NSF)} on the ``Key Concepts for Future Quantum Information Science Learners'', numerous QIS experts identified core concepts for educators to focus on \cite{NSFKeyConcepts2020}. In turn, recent pedagogical approaches have compared and contrasted these key quantum concepts to classical ones \cite{Singh2022, Schneble2025}. Despite these educational advances, there have been many questions regarding which QIS concepts should be assessed and how \cite{Meyer2024}.

In this paper, we present a new QIS assessment named the \textit{Quantum Information Science Concept Introductory Test}, or \textit{QISCIT}. As implied by the name, our concept test focuses on introductory QIS concepts that can be demonstrated without advanced mathematics or formalisms required by other quantum assessments. Hence, we anticipate that QISCIT can be used in any setting in which it would be useful to assess one's conceptual understanding of QIS, as opposed to one's procedural knowledge of QIS. Therefore, we imagine QISCIT will be used in a variety of settings, including colleges, high schools, and perhaps even lower grade levels or informal educational settings.

Furthermore, QISCIT integrates many of the recently identified key concepts, such as entanglement and decoherence, while de-emphasizing particle-based quantum mechanics concepts that many prior quantum assessments have focused on. We iteratively developed QISCIT in collaboration with 11 QIS experts to evaluate its content validity. Therefore, we believe QISCIT directly addresses prior concerns about which concepts to assess and how.

\section{Prior Quantum Assessments}

\begin{table*}[ht]
\caption{Prior quantum assessments compared to QISCIT, including prerequisite knowledge (red), quantum mechanics concepts (orange), core quantum concepts (yellow), QIS concepts (green), and whether the assessments have been validated.}
\begin{tabular}{|c|>{\columncolor{gray!30}}c|>{\columncolor{pink}}c|>{\columncolor{pink}}c|>{\columncolor{orange!30}}c|>{\columncolor{orange!30}}c|>{\columncolor{orange!30}}c|>{\columncolor{yellow!30}}c|>{\columncolor{yellow!30}}c|>{\columncolor{yellow!30}}c|>{\columncolor{green!30}}c|>{\columncolor{green!30}}c|>{\columncolor{green!30}}c|>{\columncolor{green!30}}c|>{\columncolor{blue!30}}c|}
\hline
\textbf{Assessment}                                                                          & \rotatebox[origin=l]{90}{\textbf{\# of Questions}} & \rotatebox[origin=l]{90}{\textbf{Advanced Math}} & \rotatebox[origin=l]{90}{\textbf{Dirac Notation}} & \rotatebox[origin=l]{90}{\textbf{Time Evolution}} & \rotatebox[origin=l]{90}{\textbf{Wave Functions}} & \rotatebox[origin=l]{90}{\textbf{Uncertainty Principle\hspace{0.5em}}} & \rotatebox[origin=l]{90}{\textbf{Quantum States}} & \rotatebox[origin=l]{90}{\textbf{Superposition}} & \rotatebox[origin=l]{90}{\textbf{Measurement}} & \rotatebox[origin=l]{90}{\textbf{Gates}} & \rotatebox[origin=l]{90}{\textbf{Entanglement}} & \rotatebox[origin=l]{90}{\textbf{Teleportation}} & \rotatebox[origin=l]{90}{\textbf{Decoherence}} & \rotatebox[origin=l]{90}{\textbf{{Validated}}} \\ \hline
QMCA\cite{QMCA}                                                                        & \hspace{0.5em}38\hspace{0.5em}                       & \hspace{0.5em}✓\hspace{0.5em}                           & \hspace{0.5em}✓\hspace{0.5em}                       & \hspace{0.5em}✓\hspace{0.5em}                       & \hspace{0.5em}✓\hspace{0.5em}                       & \hspace{0.5em}✓\hspace{0.5em}                              & \hspace{0.5em}✓\hspace{0.5em}                       & \hspace{0.5em}✓\hspace{0.5em}                      & \hspace{0.5em}✓\hspace{0.5em}                    &                       &               &                       & \hspace{0.5em}✓\hspace{0.5em}                    & \hspace{0.5em}✓\hspace{0.5em}                  \\ \hline
QMFPS\cite{QMFPS}                                                                                 & \hspace{0.5em}34\hspace{0.5em}                       & ✓                           & ✓                       & ✓                       & ✓                       & ✓                              & ✓                       & ✓                      & ✓                    &                       &                &                        &                      & ✓                  \\ \hline
QMAT\cite{QMAT}                                                                                  & \hspace{0.5em}14\hspace{0.5em}                       & ✓                           & ✓                       & ✓                       & ✓                       & ✓                              & ✓                       & ✓                      & ✓                    &                       &                &                        &                      & ✓                   \\ \hline
\hspace{0.5em}Hu et al.\cite{Hu2024}  & \hspace{0.5em}11\hspace{0.5em}                       & ✓                           & ✓                       &                         &                         &                                & ✓                       & ✓                      & ✓                    & \hspace{0.5em}✓\hspace{0.5em}                     & \hspace{0.5em}✓\hspace{0.5em}              &                        &                      &                   \\ \hline
Zable et al.\cite{Zable2020}  & \hspace{0.5em}10\hspace{0.5em}                       & ✓                           & ✓                       &                         &                         &                                & ✓                       & ✓                      & ✓                    & ✓                     & ✓              &                        &                      &                    \\ \hline
QMS\cite{QMS}                                                                                   & \hspace{0.5em}31\hspace{0.5em}                       & ✓                           &                         & ✓                       & ✓                       & ✓                              & ✓                       & ✓                      & ✓                    &                       &                &                        &                      & ✓                  \\ \hline
QMVI\cite{QMVI}                                                                                  & \hspace{0.5em}25\hspace{0.5em}                       & ✓                           &                         & ✓                       & ✓                       & ✓                              & ✓                       & ✓                      & ✓                    &                       &                &                        &                      & ✓                  \\ \hline
\hspace{0.5em}{Salehi et al.}\cite{Salehi2022} & \hspace{0.5em}7\hspace{0.5em}                        & ✓                           &                         &                         &                         &                                &                         & ✓                      &                      & ✓                     & ✓              & \hspace{0.5em}✓\hspace{0.5em}                      &                      &                    \\ \hline
\hspace{0.5em}DeVore and Singh\cite{DeVore2020}                             & \hspace{0.5em}8\hspace{0.5em}                        &                             & ✓                       &                         &                         &                                & ✓                       & ✓                      & ✓                    &                      &     ✓           &                        &                      &                   \\ \hline
QME\cite{QME}                                                                                   & \hspace{0.5em}20\hspace{0.5em}                       &                             &                         & ✓                       & ✓                       & ✓                              & ✓                       & ✓                      & ✓                    &                       &                &                        &                      & ✓                  \\ \hline
{QMCS}\cite{QMCS}                                                                                  & \hspace{0.5em}12\hspace{0.5em}                       &                             &                         &                         & ✓                       & ✓                              & ✓                       & ✓                       & ✓                    &                       &                &                        &                      & ✓                  \\ \hline
{QPCS}\cite{QPCS}                                                                                  & \hspace{0.5em}25\hspace{0.5em}                       &                             &                         &                         & ✓                       & ✓                              & ✓                        & ✓                       & ✓                    &                       &                &                        &                      & ✓                  \\ \hline
\hspace{0.5em}\textbf{QISCIT}\hspace{0.5em}                                                                       & \hspace{0.5em}\textbf{31}\hspace{0.5em}              & \textbf{}                   & \textbf{}               & \textbf{}               & \textbf{}               & \textbf{}                      & \textbf{✓}              & \textbf{✓}             & \textbf{✓}           & \textbf{✓}            & \textbf{✓}     & \textbf{✓}             & \textbf{✓}           & \textbf{✓}         \\ \hline
\end{tabular}
\label{tab:related_work}
\end{table*}

Numerous assessments have been developed to evaluate one's understanding of quantum concepts. Table \ref{tab:related_work} provides a comparison of these previous quantum assessments. We briefly discuss notable differences between QISCIT and these prior assessments in this section.

Most prior assessments involve quantum aspects that require advanced mathematics, Dirac notation, or both. While these prerequisites are suitable for students with strong mathematical foundations, they make these assessments largely inaccessible for students who are not experienced with linear algebra or these formalisms. Hence, this group of assessments is generally not practical for evaluating the QIS knowledge of high school students. Therefore, we designed QISCIT to not require such advanced prerequisite knowledge and to make it more practical for pre-college QIS courses. 

The majority of prior assessments have focused heavily on quantum mechanics concepts like time evolution, wave functions, and the uncertainty principle. Many include questions on all three quantum mechanics concepts. While highly applicable for quantum mechanics or physics courses, such assessments are less relevant for QIS courses that emphasize quantum information instead of physical phenomena.
Hence, we purposefully designed QISCIT to not require knowledge or understanding of such quantum mechanics concepts.

Nearly all the assessments include questions addressing quantum states, superposition, and measurement. This is not surprising given that these are foundational concepts to both quantum mechanics and QIS. As such, we carefully designed QISCIT to ensure these concepts were well covered across our questions. 

Only a few of the prior quantum assessments address the key concepts for future QIS learners, such as gates, entanglement, teleportation, and decoherence. Three of these assessments address both gates and entanglement \cite{Hu2024, Zable2020, Salehi2022}. The assessment developed by Salehi et al. \cite{Salehi2022} is the only prior assessment to the best of our knowledge to address quantum teleportation. Similarly, the QMCA \cite{QMCA} is the only prior assessment to address decoherence. In designing QISCIT, we ensured that QISCIT addressed the previously identified key QIS concepts and included questions involving gates, entanglement, teleportation, and decoherence.

Content validity (i.e., whether or not an instrument addresses what it meant to be measured) is critical for using assessments in educational and empirical settings \cite{Lopez2015}. While some of the prior assessments were validated in terms of content, several were not. To evaluate and ensure the content validity of QISCIT, we carefully employed the process of Almanasreh et al. \cite{Almanasreh2019} in our research.



















\section{Development of QISCIT}

The development of instruments should involve three steps: concept identification, item generation, and instrument formation \cite{Lynn1986}. We describe these steps for QISCIT. However, we first discuss our focus on introductory QIS students, including potential high school students.

\subsection{Audience Focus}

When developing QISCIT, we chose to focus on a broad group of students that may potentially take an introductory course on QIS, This included students from different disciplines (e.g., physics, computer science, and electrical/computer engineering). Additionally, while most introductory courses on QIS are offered to undergraduate and graduate students, we decided to specifically include high school students considering the importance of training a future quantum workforce and the European Union's goal of introducing quantum in high schools \cite{EU2024}.

Given our inclusion of high school students, we chose to develop QISCIT to not require calculus, linear algebra, or Dirac notation. We excluded calculus because it is normally taught as an advanced course in high school and is not taken by all students. We excluded linear algebra because it is rarely taught in high school, and not all college students must take it. Finally, we excluded Dirac notation, as it is a linear algebra-based notation for quantum states and operations.

In addition to avoiding advanced mathematics, we also decided to directly incorporate a glossary of key QIS terms at the beginning of QISCIT. The National Research Council's framework for K-12 science education recommends that students encounter and practice scientific terms through exercises, as opposed to rote memorization \cite{NRCScienceFramework2012}. Our glossary allowed us to create questions that required conceptual knowledge (i.e., understanding and using the relationships among concepts) instead of declarative knowledge and simply memorizing facts \cite{Miller2007}. Hence, students with an aptitude for science and making connections between concepts may answer some QISCIT questions correctly without prior QIS knowledge due to our glossary. We believe this makes QISCIT useful to instructors as a baseline measurement at the beginning of a course to measure meaningful change in students' QIS knowledge over time. 

To further support its use with
novice students, we decided to create multiple-choice questions for QISCIT, employing either four or five choices per question. This facilitates students with little prior QIS knowledge and experience to still deduce correct answers if they understand the concepts, as opposed to thinking of open-ended responses. It also makes QISCIT efficient and scalable, with the ability to automatically score answers and easily administer the instrument to large groups. The multiple-choice format also allowed us to design distractors (i.e., incorrect choices) that reflect potential student errors and can help diagnose misconceptions. 

\subsection{Concept Identification}

To identify which QIS concepts to assess with QISCIT, we focused on the key concepts identified in the prior NSF workshop report \cite{NSFKeyConcepts2020}, which include: the definition and scope of QIS, quantum states, quantum measurement, qubits, entanglement, coherence and decoherence, quantum gates/computing, quantum communication, and quantum sensing. A recent survey of introductory QIS courses by Meyer et al. \cite{Meyer2024} reinforces the importance of most of these topics except for quantum sensing, which was covered in only 22\% of QIS courses. We decided not to include quantum sensing as a concept when developing QISCIT, as it is not usually covered in computer science QIS courses \cite{Meyer2024}, nor usually introduced using gates. 

\subsection{Item Generation}

The assessment triangle is a framework presented by the National Research Council for designing and evaluating educational assessments in a systematic way \cite{NRCEducationalAssessment2001}. It consists of three connected aspects: 1) \textit{cognition}: a model or theory of what students should learn; 2) \textit{observation}: an assessment that elicits evidence of their knowledge and understanding; and 3) \textit{interpretation}: inferences about what each student knows or understands based on that evidence. We addressed the cognition aspect through the identification of which concepts QISCIT would cover. We addressed the observation and interpretation aspects of the assessment triangle when generating items.

As mentioned above, we focused on creating questions that required conceptual knowledge, rather than declarative knowledge, to answer correctly. In turn, these questions elicit evidence of a student's understanding of the concepts, as opposed to simple rote memorization of the concepts. For example, consider the following question:

{\sffamily
\vspace{1em}
{\par \noindent Two qubits are entangled, and then the first qubit is measured. What happened to the entanglement between the two qubits?}
\begin{itemize}[noitemsep]
    \item[✓] The entanglement decreased to zero \vspace{0.5em}
    \item[O] The entanglement depends on the result of the measurement \vspace{0.5em}
    \item[O] The entanglement increased \vspace{0.5em}
    \item[O] The entanglement remained the same
\end{itemize}
}

This question tests the student's knowledge of quantum states, quantum measurement, qubits, and entanglement. First, the student must understand what qubits are and what it means for them to be entangled. The student must then understand what it means for the first qubit to be measured. The correct answer of the entanglement decreasing to zero requires the student to recognize that the measurement causes the quantum state shared by the two entangled qubits to collapse. 

As we created questions for QISCIT, we also carefully considered the distractors to provide inferences about what the student does not understand and what misconceptions they may have. Consider the example question again. If the student indicates that the entanglement depends on the result of the measurement, they may not understand quantum measurement, or, more likely, they confused the entanglement with the state of the second qubit depending on the result of the measurement. If the student indicates that entanglement increased, the student has a clear misunderstanding about the relationship between measurement and entanglement. Finally, if the student indicates that entanglement remained the same, the student most likely does not understand that there is a relationship between measurement and entanglement. 
	



\subsection{Instrument Formation}

After drafting and generating initial questions for QISCIT, we followed established guidelines for creating high-quality, multiple-choice questions, such as ensuring the stem asks a clear question that can be answered without the choices and all choices are approximately equal in length and detail \cite{Tarrant2012}. After refining our items, our initial version of QISCIT included 22 questions. Due to the entwined nature of the QIS concepts, most questions addressed multiple concepts: quantum states (21 questions), quantum measurement (12 questions), qubits (21 questions), entanglement (12 questions), coherence and decoherence (3 questions), quantum gates/computing (12 questions), and quantum communication (3 questions). The questions also addressed several subconcepts, including basis states, superposition, quantum gates like the Hadamard (H) gate, phase gates (S and T),
Pauli X gate, controlled-not (CNOT) gate, and quantum teleportation. The supplementary material includes concept mappings for each question.

\section{Content Validation of QISCIT}

Once the initial version of QISCIT was drafted, we employed the process recommended by Almanasreh et al. \cite{Almanasreh2019} to evaluate the content validity of the instrument. This process involved recruiting a group of experts with QIS education experience to provide feedback on the instrument through two rounds of evaluation. This group included 11 QIS experts from a broad set of institutions and disciplines, including physics (5 experts), computer science (3 experts), electrical/computer engineering (2 experts), and chemistry (1 expert).

In both rounds, the experts separately and remotely reviewed the current instrument, including the glossary and each individual question. For every item, each expert was required to rate the content relevance of the item from 1 (``Not Relevant'') to 4 (``Extremely Relevant''). The content validity index (CVI) of each item was calculated by counting the number of experts who rated the item as 3 (``Relevant'') or 4 (``Extremely Relevant'') and dividing by the total number of experts. As recommended by Almanasreh et al. \cite{Almanasreh2019}, we employed the 0.78 item-level CVI threshold suggested by Polit et al. \cite{Polit2007} for accepting or rejecting individual items. We also calculated the average CVI across the items, which represents the overall content validity of the instrument, to ensure it was acceptable. 

In addition to rating each item, the experts were required to answer each question. The experts also had the opportunity to provide open-ended feedback on each item, including the glossary. In addition to considering the CVI of each item, we examined the experts' accuracy and feedback to identify any potentially confusing items, which were then either revised or removed for clarity. Below, we discuss the results of each round of the experts' feedback.

\subsection{First Round of Expert Feedback}

After receiving the first round of expert feedback, the average CVI across the items was high (0.85), and the average accuracy was reasonable (87\%). However, two questions were dropped based on their CVI, accuracy scores, and/or open-ended feedback. One focused on entanglement (CVI = 0.45, accuracy = 45\%), and one focused on teleportation (CVI = 0.82, accuracy = 73\%).

Of the remaining 20 questions, 18 were revised based on suggestions from the experts on how to improve their relevance and/or clarity. Of those, 15 question stems were revised, ranging from complete rewordings to adding a single word to including a circuit diagram. The inclusion of circuit diagrams was an extremely beneficial suggestion from one of our experts that improved the clarity of our more-complex questions without requiring advanced mathematics or Dirac notation. The revised questions also involved 13 sets of revised choices, ranging from completely revised sets (e.g., all four choices were revised) to single choices being revised (often to replace a poor distractor identified by the experts).

The experts also suggested concepts that they thought should be emphasized, such as teleportation, or added to QISCIT, including Bell states, Pauli Y and Z gates, unknown gates, and unknown states. We developed 14 new questions to address these concepts and the feedback. 

The glossary's relevance was also rated by the experts and received a CVI score of 1, indicating strong content validity. We revised the definitions of several terms based on the experts' feedback, including qubit, measurement, Bloch vector, quantum gate, each of the individual quantum gates, entanglement, and decoherence. We also added new terms to the glossary based on the concepts emphasized by the experts, including the Y gate, Z gate, quantum circuit, Bell state, and quantum teleportation.

Our revisions after the first round of expert feedback resulted in an updated version of QISCIT that included 34 questions, which is documented in the supplementary material. Again, each question addressed multiple QIS concepts: quantum states (34 questions), quantum measurement (20 questions), qubits (34 questions), entanglement (15 questions), coherence and decoherence (3 questions), quantum gates/computing (24 questions), and quantum communication (6 questions). We sent this updated version of QISCIT to our 11 QIS experts for the second round of feedback.

\subsection{Second Round of Expert Feedback}

The average CVI across the items was very high (0.95), indicating strong content validity for QISCIT overall, and the average accuracy increased (89\%). However, three items needed to be dropped based on their CVI, accuracy scores, and/or open-ended feedback. One item had been previously revised after the first round of expert feedback, but its CVI (0.73) and accuracy (64\%) remained low after the second round of feedback. Two other items were newly added for the second iteration of QISCIT but were ultimately dropped. While they had acceptable CVI scores (CVI = 0.82 for both), their accuracy scores were low (64\% and 73\%), and the experts had concerns about how the questions were worded. After removing these three questions, the average CVI across the remaining 31 items was even higher (0.97), and the accuracy further increased (91\%). The glossary's content validity was again very strong in the second round (CVI = 1). 

Three of the remaining items were revised based on suggestions from the experts on how to improve their relevance and/or clarity. One question stem was slightly reworded based on the experts' feedback. Similarly, one distractor for one of the questions was revised based on the experts' comments. The biggest revision involved changing the starting states for two qubits in a question from state 1 to state 0, which also changed the correct choice. We made this change because one expert explicitly commented about never starting in state 1 and because the accuracy was low (64\%), likely due to the uncommon starting premise. 

\section{Final Version of QISCIT}

The final version of QISCIT includes the glossary and the remaining 31 multiple-choice questions randomized for assessments. Each question addresses multiple key QIS concepts, which provides multiple observations and inferences for each concept: quantum states (31 questions), quantum measurement (19 questions), qubits (31 questions), entanglement (13 questions), coherence and decoherence (3 questions), quantum gates/computing (23 questions), and quantum communication (6 questions). The questions also cover numerous subconcepts, including basis states, superposition, Bell states, quantum gates (like H, S, T, X, Y, Z, and CNOT gates), quantum circuits, quantum teleportation, unknown gates, and unknown states.

Based on our second round results, QISCIT has strong overall content validity (CVI = 0.97). While CVI provides a crucial data point about whether a question is useful to include, the accuracy of the experts in answering the question provides more nuanced information. On the one hand, low accuracy scores in the first round often indicated poor or confusing wordings, which were corroborated by low CVI scores and/or concerns expressed by the experts. By revising or removing these questions, our overall accuracy went from 87\% in the first round to 89\% in the second round to 91\% in the final version (obtained by dropping problematic questions from the second round). However, note that the accuracy never reached 100\% despite the fact that the assessment was taken by QIS experts with quantum education experience. One reason for this is that some of the problems are quite challenging (see Example Question \#3 below), such that even an expert forced to reason through them could unwittingly make a mistake. We do not believe this is a fundamental problem, as it shows that even a conceptual, non-mathematical QIS assessment can encompass vital but challenging topics. 

A second reason for this imperfect accuracy is that it seems to reflect a fundamental reality about QIS conceptual understanding today. QIS is a challenging field with many opportunities for misconceptions, given the conflicts between our classical day-to-day life and the quantum reality. Experts validating this assessment came from many backgrounds---physics, computer science, electrical/computer engineering, and chemistry---with various perspectives, practices, and expertise in QIS. During the content validation process, we observed notable differences in accuracy scores and commentary among the experts, usually distinguished by their respective fields. For example, our experts from physics were less likely to make mistakes pertaining to quantum states while our experts from computer science were less likely to make mistakes pertaining to quantum computing and gates. We believe this does not indicate an issue with our assessment but rather highlights the highly interdisciplinary nature of QIS and the importance of having an assessment like QISCIT. The first step in improving QIS education is to determine areas of imperfect understanding, and we believe our results indicate that QISCIT will do this effectively.

Furthermore, we expect QIS instructors will find QISCIT useful in a variety of ways. As previously mentioned, we think that instructors can administer QISCIT at the beginning of a course to serve as a baseline measurement. Our detailed concept mappings in the supplementary material can be used to filter out irrelevant questions due to uncovered topics (e.g., decoherence) or to hand-select highly relevant questions for particular topics (e.g., Bell states). Instructors can also use QISCIT to look at how their students' knowledge of various, important QIS concepts changes throughout a course. Also, instructors who want to challenge their students can use QISCIT's questions without the choices as free-response items since we followed best practices for designing multiple-choice questions.

\section{Example Questions from QISCIT}

The full version of QISCIT is available in the supplementary material. Here, we give a few examples of questions that best illustrate the design principles and content of QISCIT.


\subsection{Example Question \#1}

{\sffamily
\vspace{1em}
{\par \noindent A qubit starts in an unknown superposition state. An H gate is applied, followed by a measurement. How can the qubit be returned to its original unknown state?}
\begin{center}
    \includegraphics[height=3em]{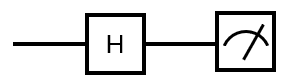}
\end{center}
\begin{itemize}[noitemsep]
    \item[O] Apply another H gate and then measure again \vspace{0.5em}
    \item[O] Apply another H gate but do not measure again \vspace{0.5em}
    \item[O] Measure the qubit again and then apply another H gate \vspace{0.5em}
    \item[✓] It is not possible to recover the initial unknown state
\end{itemize}
}

Similar to the example shown earlier, this problem touches on many key concepts at the same time: quantum states, quantum measurement, qubits, and quantum gates/computing. The correct answer highlights a fundamental property of quantum mechanics---the fact that measurement causes state collapse and is irreversible---which commonly leads to misconceptions. The distractor options are designed to see where a student's potential misconception occurs, as each represents an incorrect line of logic for how one might reverse the measurement.

There are two important changes that occurred to this question during the iteration process. First, rather than giving solely a text-based question stem, we have added a quantum circuit diagram wherever possible. This was directly based on the first round of expert feedback, prior to which no quantum circuits were included. Second, the problem was modified to make the circuit explicit. In the first round, a similar question simply asked ``{\sffamily When are measurements not reversible?}'' This led to potential confusion. For example, if measuring from the initial state 0, is the measurement ``reversible'' given that no collapse occurs? By replacing this abstract question with an explicit circuit, we now test the same concept but without such sources for confusion. The concrete circuit also allows for more-explicit distractor options that test for potential misconceptions, such as confusing the reversibility of unitary gates with the irreversibility of measurements.

\subsection{Example Question \#2}

{\sffamily
\vspace{1em}
{\par \noindent Two qubits each start in state 0. An H gate is applied to the first qubit. Two CNOT gates are then applied, using the first qubit as the control and the second qubit as the target. If the second qubit is then measured, what is the probability it will return state 0?}
\begin{center}
    \includegraphics[height=5em]{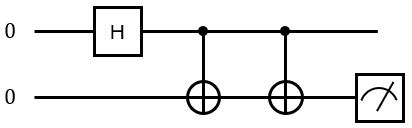}
\end{center}
\begin{itemize}[noitemsep]
    \item[O] 0\% \vspace{0.5em}
    \item[O] 25\% \vspace{0.5em}
    \item[O] 50\% \vspace{0.5em}
    \item[O] 75\% \vspace{0.5em}
    \item[✓] 100\%
\end{itemize}
}

This question illustrates two key design principles of QISCIT: a) addressing multiple concepts at the same time (quantum states, quantum measurement, qubits, entanglement, and quantum gates/computing), and b) the use of specific quantum circuits to test conceptual understanding. Furthermore, by tweaking a canonical Bell state circuit with the addition of a second CNOT, this question has the potential to detect ``quantum memorizers,'' who might simply have learned a handful of standard circuits without any deeper understanding and will try to answer accordingly. Instead, the correct answer is most easily obtained by realizing a different concept---the reversibility of unitary quantum gates. When consecutively applied twice to the same control and target qubits, the CNOT gates will either flip the target twice (i.e., a double NOT) or will not flip the target in either application. Of course, the correct answer can be deduced by either realizing that the CNOT gate applied twice squares to the identity matrix or by directly tracing the quantum circuit using two-qubit wave functions. However, like all problems in QISCIT, the correct answer can be obtained without explicit use of linear algebra, emphasizing the conceptual aspects of QIS.

\subsection{Example Question \#3}

{\sffamily
\vspace{1em}
{\par \noindent The following quantum circuits produce orthogonal Bell states. What common single-qubit gate could have been applied as the unknown gate?}
\begin{center}
    \includegraphics[height=5em]{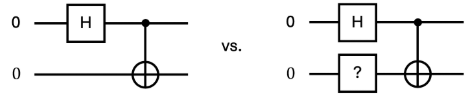}
\end{center}
\begin{itemize}[noitemsep]
    \item[O] An H gate \vspace{0.5em}
    \item[O] An S gate \vspace{0.5em}
    \item[✓] A Y gate \vspace{0.5em}
    \item[O] A Z gate
\end{itemize}
}

In addition to demonstrating similar design principles as the previous two examples, this item demonstrates another question format common to QISCIT, in which students are asked to determine an unknown quantum gate. This allows us to take abstract ideas and encode them in explicit circuits, where each available answer is a proxy for a potential misconception. This particular problem is one of the hardest in QISCIT based on expert feedback (``This one was tricky!''). The simplest route to a correct answer involves understanding the meaning of orthogonality of quantum states and realizing that orthogonality is conserved by unitary operations.
Therefore the correct gate---the Y gate---is the one that flips the second qubit from state 0 to state 1, producing orthogonal inputs to the CNOT gate. The remaining options either prepare a second superposition (the H gate) or leave the initial state 0 unchanged (the S and Z gates), neither of which is orthogonal to the original state. Crucially, this logic is fully achievable without any explicit calculations via linear algebra. Indeed, solving the problem using linear algebra is significantly more challenging. As a final note, we chose the correct answer to be the Y gate, rather than the X gate, in order to balance our coverage of Pauli gates across QISCIT's questions. This maximizes the chances of identifying conceptual issues with a minimum number of questions.

\section{Conclusion}

We have created a new assessment instrument that we call the Quantum Information Science Concept Introductory Test (QISCIT). While most prior quantum assessments require advanced mathematics and emphasize quantum mechanics concepts, we designed QISCIT for a broad range of QIS courses, including those for high school students, while focusing on key QIS concepts recently identified by numerous experts. Our development process followed the assessment triangle framework, which generated questions requiring conceptual knowledge rather than rote declarative knowledge. With the help of 11 QIS experts, we iteratively improved upon QISCIT and validated its content. The resulting 31 questions cover important quantum concepts, such as quantum states, quantum measurement, qubits, entanglement, coherence and decoherence, quantum gates/computing, and quantum communication. 

We have discussed how QIS instructors can use QISCIT in a variety of ways for their courses. In the supplementary material, we have shared all our results and all three iterations of our assessment, including the current validated version of QISCIT. The linked Open Science Framework (OSF) serves as a living repository of our work on QISCIT, as we plan to continue iterating and refining the assessment for the benefit of QIS instructors and students. We have already administered QISCIT in multiple undergraduate QIS courses and are working to evaluate the reliability of QISCIT.

\section{Supplementary Material}

In the supplementary material, we provide PDF, Word, and Qualtrics formats for each of the three versions of QISCIT, including the two earlier iterations and the current validated version. In addition to the various formats, we also include the concept mappings, average CVI score, and average accuracy for each question. 

Please click the following link to access our living Open Science Framework (OSF) repository for QISCIT: \href{https://osf.io/b5ezc/?view_only=0d78ce3d33d14aa3bd3b8f5d54d37de6}{https://osf.io/qiscit/viewonly}.

\begin{acknowledgments}

We are grateful for the hard work and detailed feedback from our QIS experts. This material is based upon work supported by the National Science Foundation under Grant Nos. 2302816, 2302817, 2302818, and 2516450. Any opinions, findings, and conclusions or recommendations expressed in this material are those of the author(s) and do not necessarily reflect the views of the National Science Foundation. 

\end{acknowledgments}





\end{document}